# Exact Requirements Engineering for Developing Business Process Models

Masoud Nosrati
Department of Computer Engineering
Kermanshah Branch, Islamic Azad University
Kermanshah, Iran
minibigs_m@yahoo.co.uk

*Abstract*—Process modeling is a suitable tool for improving the business processes. Successful process modeling strongly depends on correct requirements engineering. In this paper, we proposed a combination approach for requirements elicitation for developing business models. To do this, BORE (Business-Oriented Requirements Engineering) method is utilized as the base of our work and it is enriched by the important features of the BDD (Business-driven development) method, in order to make the proposed approach appropriate for modeling the more complex processes. As the main result, our method eventuates in exact requirements elicitation that adapts the customers' needs. Also, it let us avoid any rework in the modeling of process. In this paper, we conduct a case study for the paper submission and publication system of a journal. The results of this study not only give a good experience of real world application of proposed approach on a web-based system, also it approves the proficiency of this approach for modeling the complex systems with many sub-processes and complicated relationships.

## I. INTRODUCTION

Initial step and probably the most difficult part of designing an information system is to decide what to build. Information systems let the corporations to perform their process automatically, regarding the logic and complex business rules for achieving the future goals [1,2]. When implementing an information system, insufficient requirements analysis in creating the business model causes unsatisfactory results that cannot fulfill the requirements of the business process. Process models have been a popular tool for developing business in the past decades. Modeling the business processes have been enhanced from different perspectives, like optimizing the work process, developing organization practices to fulfill the quality standards, automating work, and developing IT systems. Successful modeling of a business process strongly depends on the quality of the requirements engineering of the process. Nuseibeh and Easterbrook [3] define the requirements engineering as the process of identifying stakeholders and their needs, and documenting these in a form that is amenable to analysis, communication, and subsequent implementation. Przybyłek [4] outlines two types of essential and accidental difficulties and the challenges of requirement engineering as follows:

1. Essential difficulties:
   - Understanding what the customers need, while they often have a vague picture of their requirements [5,6]
   - Effective communication among the stakeholders due to the gap between the business process and the system domains [7]
   - Adaptation to the frequent and arbitrary changes of requirements [5]

2. Accidental difficulties:
   - Inadequate requirements elicitation practices, when the customers cannot articulate requirements or they are not involved enough in the requirements engineering process.
   - Rework and miscommunications due to many notations that are utilized by different stakeholders, like BPMN and UML [8]
   - Deficiencies in backward traceability make it difficult to keep consistency between documentation and the underlying information system [9,10]

3. Challenges:
   - Business analysis must precede requirements elicitation in order to obtain a deep enough understanding about the organization.
   - Using the same notation through the whole project enables all stakeholders to share the same work products.
   - Documentations must link business processes to results of an analysis, design and implementation in an explicit and traceable manner.

In the proposed method, two methods of requirements engineering, namely "Business-Oriented Requirements Engineering (BORE)" [4] and "Business-driven development (BDD)" [5] are utilized for coping with the aforementioned challenges. This approach allows for extracting the process requirements from business process models and enables traceability between business processes and the corresponding system requirements for ensuring that the system requirements meet real business needs. One of the main goals of BORE is to precede the requirements engineering parallel to the system modeling, which causes fast progress of the work. On the other hand, BDD offers some tips for making the information system more consistent. It provides a more extensive definition of requirements engineering by considering the possible future changes of the requirements as well as hierarchy of requirements and the integrity of the information system. Modifying the BORE method with some tips of the BDD eventuates in the exact still flexible requirements engineering.

The rest of this paper is organized as follows: second section casts a look at BORE and the tips of BDD in order to delineate the proposed method. Third section is dedicated to a case study of applying the proposed method on the journal publication process. Related researches are abstracted in fourth section. Finally, conclusion is placed in fifth section.

## II. METHODOLOGY

The base method of our approach is BORE in which the main objectives are [4]:
1. supporting requirements elicitation under conditions of uncertainty about client needs
2. makeing that the system requirements are in alignment with and provide support for the underlying business processes

This method is designed for the organizations that is accessible, and willing to undergo a process of innovation.

Accuracy and preciseness of obtained information plays the most important role in requirements elicitation. Due to this, choosing the suitable data collection methods is important. Data collection methods in BORE include semi-structures and unstructured interviews, apprenticing, workshops and scrutinizing the documents.

Improving the business process and implementing the information system are mutually dependent. Since the customers have not a clear picture of their needs, the data elicitation methods must be appropriate enough to eventuate in proper information. Also, the analysis of the business process and requirements elicitation is composed of some overlapping functions that accelerate the parallel preceding of them.

As shown in figure 1, BORE consists of three steps. The aim of the first step is to construct the As-Is model, which is a diagram that shows the details of current roles, relationships, etc. As-Is diagram helps to understand the organization of the business process. For creating this diagram, gathering the large amounts of information is needed. It can be performed via aforementioned data collection methods. Then, scrutinizing the obtained information to specify the exact roles, relationships, and other entities in the system is indispensible. The output of this step is the As-Is diagram. This diagram should be validated by the stakeholders to ensure that it is absolutely adaptable to the current state of the organization. Sometimes, it needs to some revision to achieve the final diagram.

The goal of the second step is to improve the process through automation. Business process is composed of some sub-processes. The costs and benefits of implementing the information system for each sub-process should be calculated based on the workload and the domain of the facilities for implementing it. Those sub-processes that are cost effective are prioritized for automation. According to BDD, capabilities and limitations of the project should be considered as an important factor during the analysis of the business process. In this step, holding workshops will be beneficial, because it helps to the convergence of the ideas and achieving fast consensus. The output of this step is the draft of To-Be model, which shows the general goals of the automation. In To-Be model, all the sub processes are labeled by *A*, *S* or *M*. The label *A* means that the sub-process should be performed automatically in the final information system. By *S*, the sub-process is executed with the support of the human agent, while *M* shows that it is performed manually. The ideal goal of the automation is to perform all the tasks automatically, but it is almost impossible in all the business processes. Some issues like human discipline, variability of the environment, legal responsibilities, authority level, limitations of the system, and the trade-off between the costs and benefits are the main barriers for this goal [11,12].

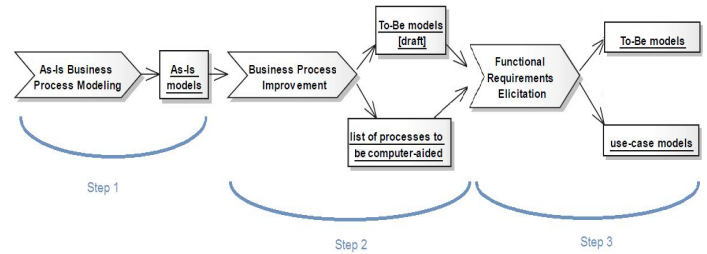

Fig. 1. Three steps of BORE [4]

Third step is to elicit the functional requirements. Holding the workshops is still a good way for deciding about the requirements. In this step, the final To-Be models are constructed for the process and the use case diagrams are drawn. Przybyłek [4] believes that there is not an effective way to algorithmically perform the functional requirements elicitation, and it highly depends on the experience of the designers.

According to BDD, there are different approaches for generating the final To-Be from the current As-Is diagram, as shown in figure 2. Based on the first approach, the improvements of the organization is the consequence of strategic developments. Some examples are: general policies of the business, the level of automation, customers' share from automation, etc. Capabilities and limitations should be considered, too. Second approach is the organizational improvement. Changing in the relationships and the capabilities of the organization, and changing the roles are some examples of this approach. Third approach is related to the business process, which is consisted of the workflows, regulations and the logic of the business and the consequential sub-processes. Based on this approach, improvements mostly affect the relationship of end-users. The last approach is the system and information level, in which the details of improvements via implementation are investigated, like integrating the information system, etc. Figure 2 also shows the hierarchy of the approaches. In other words, strategic approach focuses on the most general issues, while the system and information level approach concentrates on the minor sub-processes.

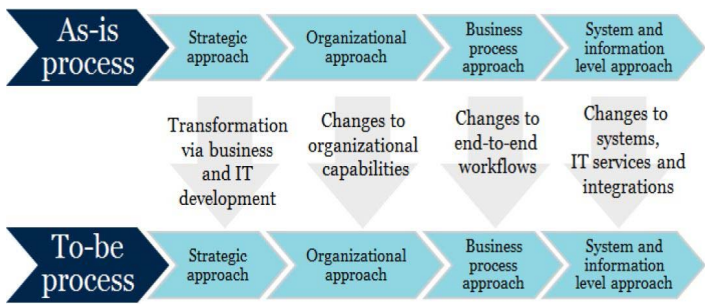

Fig. 2. Different perspectives of process modeling [3]

III. CASE STUDY

The process of journal paper submission system is selected for the case study. The selected journal is one of the scientific journals, which is published by Islamic Azad University, Sanandaj Branch. All the steps of publishing the papers are handled manually. The work process is shown in figure 3. First, the corresponding author submits the paper to the email of editorial office. Secretary does the initial check for the format of document, supplementary materials, correctness of submission, and all other issues that might be needed by editor for the evaluation. If the paper is confirmed, the secretary assigns a unique number to the submission and hands it over the editor-in-chief, and send an acknowledge email to the author. Editor-in-chief should decide about how to evaluate the paper. There are three choices: he can evaluate the paper himself, or ask one of the editorial board members to evaluate the paper, or send the paper to some reviewers and ask them to provide some comments about it, in order to help him evaluate the paper. Also editorial board members may decide to send the paper to the reviewers, and then prepare a letter for the editor-in-chief about their final decision. After all these steps, editor-in-chief will make the final decision. The paper may be accepted, rejected, or found needing to revision. If a paper needs to revision, the author is asked to amend some parts of paper according to what the editor asks him, in order to make the paper appropriate for publishing in the journal. Editor-in-chief then decides the revised paper to be accepted or not. Accepted paper will be sent for typesetting. Typesetter asks the author to submit the final version of the paper with approved authors' details like emails, affiliations, etc. Then, he prepares it according to the template of the journal. As soon as the formatting is finished, typesetter sends a copy of the paper to author for the final check and modification if needed. By the confirmation of author, the paper will be placed in the publishing queue.

The information about the workflow of the process is gathered from three sources: first, we had some interviews with secretary and editor-in-chief of the journal to make a top-down picture of the whole process. Then, an open end questionnaire was created to ask about the vague details of this process. Finally, we gathered some useful information via browsing the related websites. Figure 4 shows the As-Is model of the process.

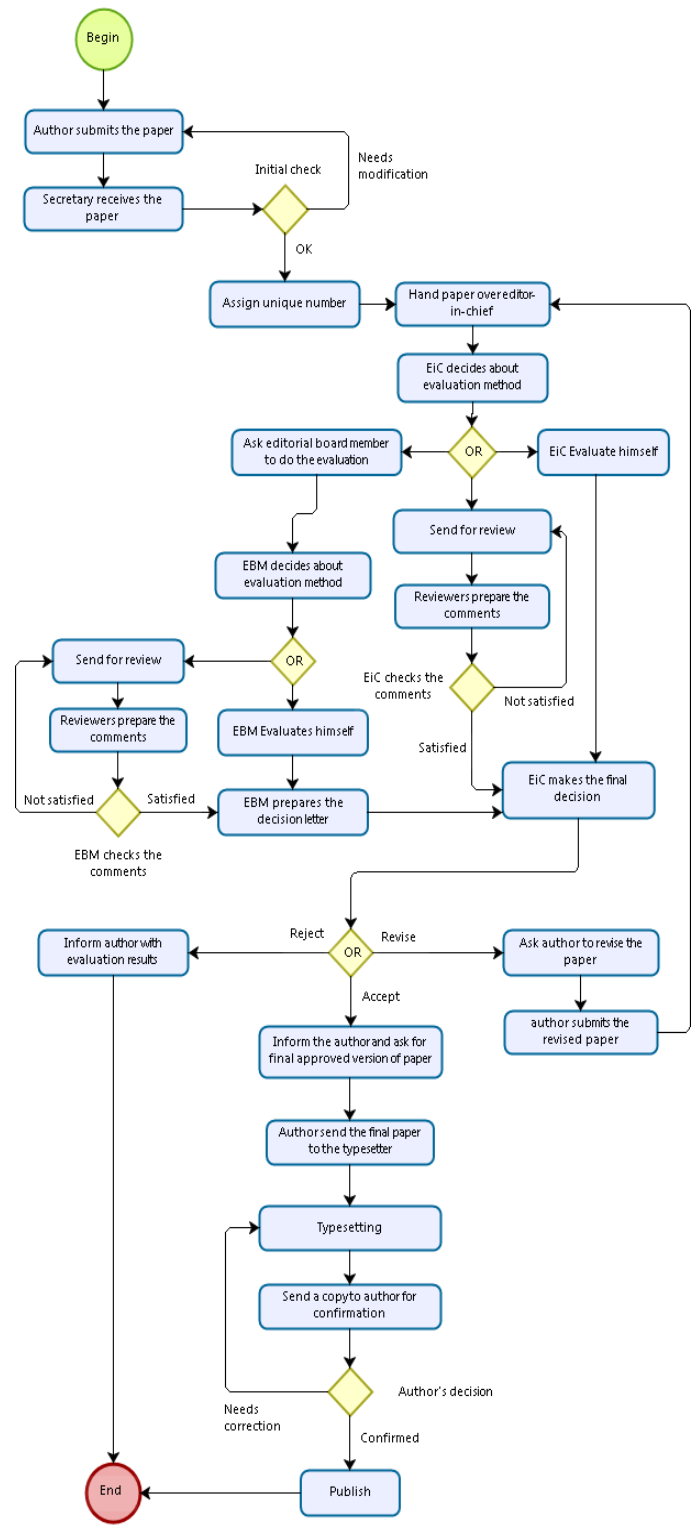

Fig. 3. The followchart of paper submission system of the journal

3th International Conference on Web Research

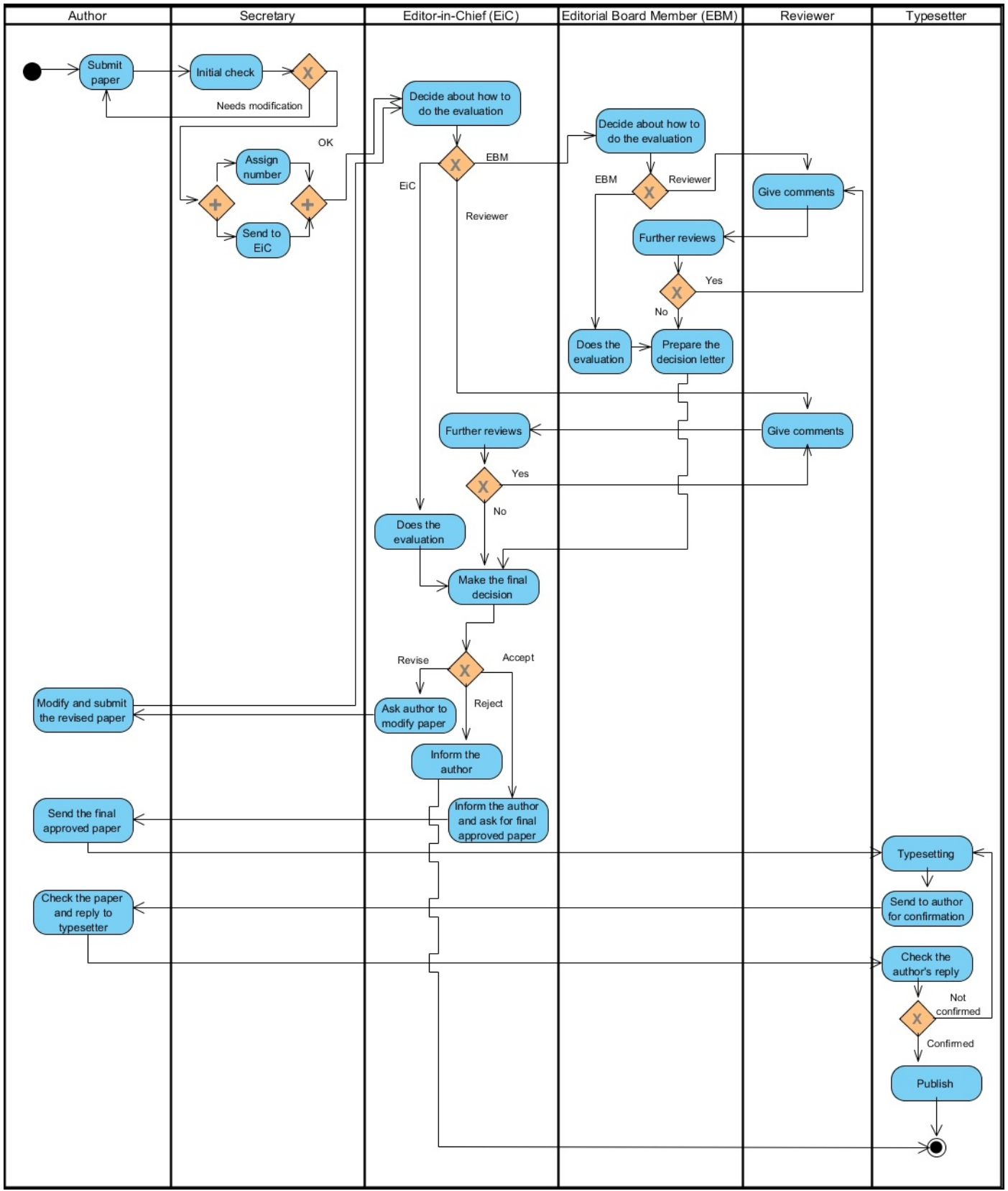

Fig. 4. As-Is diagram for the paper submission system of the journal

Figure 4 shows the As-Is diagram of the submission system that contains some sub-processes that we should make decision about them. Also, this model is validated via workshop with the stakeholders. Some minor revisions were needed to achieve the final diagram. In order to make decision for computerizing the tasks, benefits and costs of the task should be taken into account. In other words, it should be determined that computerizing the task is how much beneficial and how much cost it imposes to implementation. Those tasks who are accompanied with high benefits and low costs are prioritized for computerization. Table 1 shows the list of the tasks in the As-Is model with their benefits and costs that are marked by high, medium and low. Some of the repeated tasks are omitted because of their similarity. The last column of the table shows the decision for the task. It can be A (Automatic), S (Supported), or M (Manual). The label A means that the task will be completely computerized in the To-Be model, and S means that it will be partially computerized with the support of the human agent. The label M indicates that the task remains manual, and it is not computerized.

The next step is to create the To-Be diagram based on the decisions that we made in the table 1. To-Be model is almost the same of As-Is, with marked decisions. Figure 5 shows the To-Be diagram of the submission system. As it is shown, the tasks are shown by different colors. Green color means that the task should be performed automatically in the implemented final information system, while yellow means that the task is executed with the support of a human agent. Those tasks with blue color are still performed manually in the final system.

Validation of the To-Be model was checked through workshop with stakeholders. All the people who are working in this system confirmed that the To-Be model is absolutely adapted to their functional needs.

After the validation, it can be claimed that the proposed method for elicitation of requirements was successful, because we could follow the modeling of system, in parallel with requirements engineering. Besides, the extracted To-Be model was highly adapted to the real needs of stakeholders. Also, we avoid any rework during the modeling and requirements engineering processes, due to utilizing BDD lessons. As the last notion, the power of the proposed method was latent in the extent and depth of the information that was gathered in initial steps of the work.

From a hierarchical point of view, the submission system of the journal publishing process has three major processes: "paper submission", "evaluation" and "typesetting and publishing". Figure 6 shows the use case diagram of the system with all the five actors and three main processes. It is clear that the sub-processes of Editor-in-chief is more than other actors, because of its key role in the journal publishing process.

TABLE I. SUB-PROCESSES OF THE SUBMISSION SYSTEM WITH DECISIONS FOR COMPUTERIZING THEM

| No | Sub-process | Benefit | Cost | Decision |
|---|---|---|---|---|
| 1 | Submit the paper (Author) | High | Low | A |
| 2 | Initial check (Secretary) | Low | Low | S |
| 3 | Assign unique paper number (Secretary) | High | Low | A |
| 4 | Send the paper to EiC (Secretary) | High | Low | A |
| 5 | Decide how to evaluate the paper (EiC) | Medium | High | M |
| 6 | Decide how to evaluate the paper (EBM) | Medium | High | M |
| 7 | Give comments (Reviewer) | High | Low | A |
| 8 | Decide for further review (EiC and EBM) | Low | Medium | M |
| 9 | Evaluate the paper (EBM) | Low | High | M |
| 10 | Prepare the decision letter (EBM) | Low | Low | M |
| 11 | Evaluate the paper (EiC) | Low | High | M |
| 12 | Make the final decision (EiC) | High | High | S |
| 13 | Reject the paper and inform the author (EiC) | High | Low | A |
| 14 | Ask author to send the revised paper (EiC) | High | Medium | S |
| 15 | Accept the paper and ask author for final version of paper (EiC) | High | Medium | S |
| 16 | Submit the revised paper (Author) | High | Low | S |
| 17 | Submit the final version of paper (Author) | High | Low | S |
| 18 | Typesetting (Typesetter) | Low | High | M |
| 19 | Send the paper after typesetting for author's confirmation (Typesetter) | High | Low | A |
| 20 | Reply to the typesetter (Author) | High | Low | S |
| 21 | Check the author's reply | Low | Medium | M |
| 22 | Publish | Low | High | M |

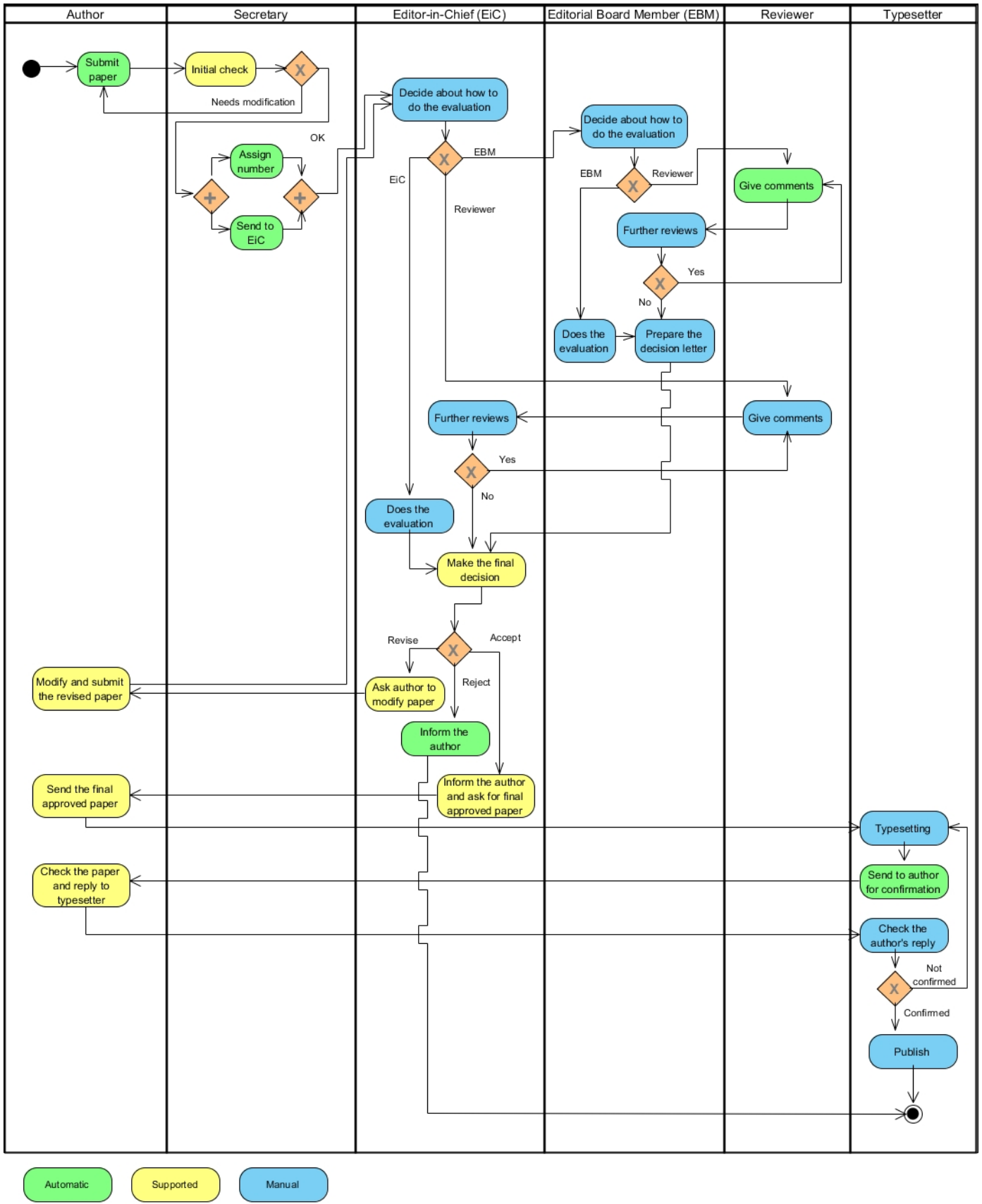

Fig. 5. To-Be diagram for the paper submission system of the journal

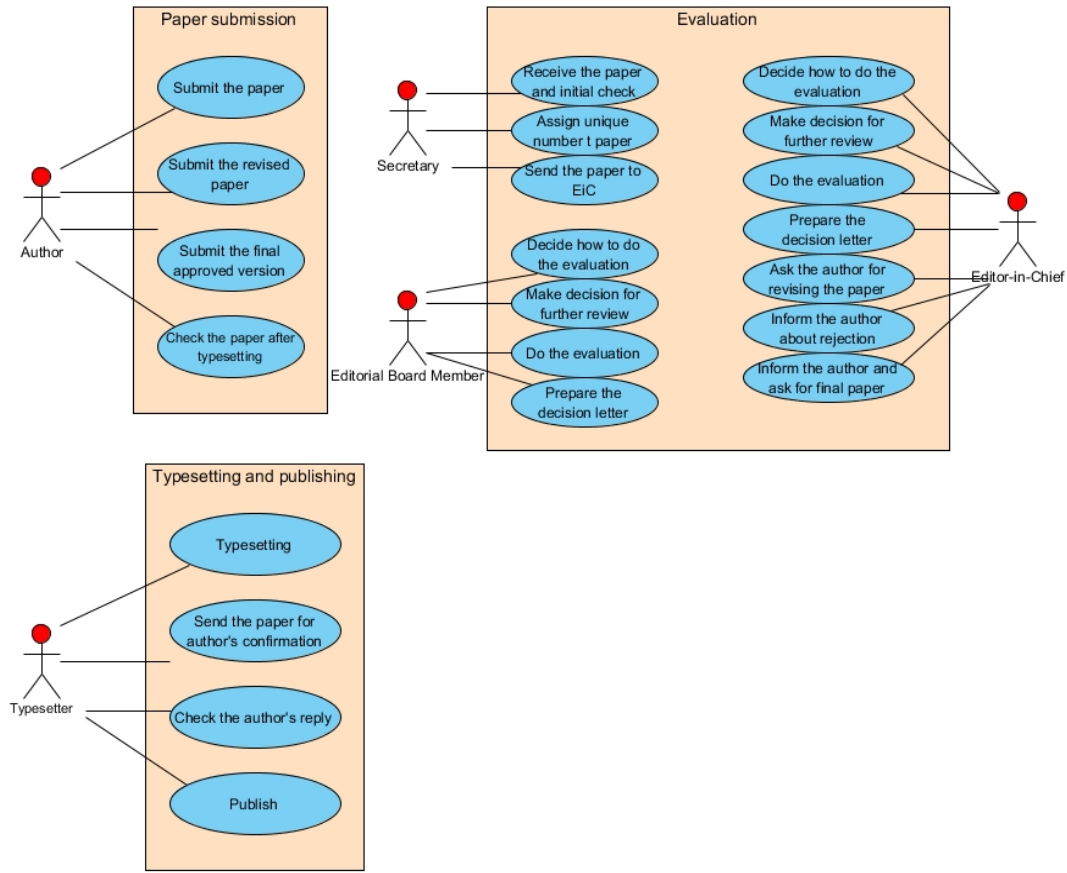

Fig. 6. Use case diagram of submission system

## IV. RELATED WORK

Online business ranging from small retail to huge cloud service providing as well as other types of services became the most popular type of business in recent years [13-17]. Gaining from different techniques in IT opened new doors for fast development of traditional business tasks [18,19]. Process modeling was in the limelight of many previous studies. Business Process Management (BPM) provides concepts, methods, and techniques for enhancing the company's business processes [20]. One of the BPM related methods the Business-driven development method (BDD) [21] combines business process and IT development. The main goal of BDD is to develop information systems that directly satisfy business requirements and functional needs [21]. There are many requirements engineering methods that make it challenging to select the most suitable technique for the project context [22]. Furthermore, the case study organization has applied several requirements modeling methods, such as UML modeling and use cases, business rules and non-functional requirements. Use cases are widely used to model interaction between user and system [23] and a business rule is a statement that defines or restricts some aspect of the business [24]. On the other hand, non-functional requirements are often needed [25,26]. Glinz [25] defines that a non-functional requirement is an attribute of or a constraint on a system, and that these attributes may include both performance requirements and specific quality requirements.

Štolfa and Vondrák [27] claimed that there are repeatable situations during transition between modeling the business process and requirement elicitation. Also, they described three patterns that can be applied to support the transition between business processes modeling as well as other phases of software development.

In recent years, the importance of enterprise architecture (EA) has increased, as the number of techniques and complexity of the solutions has risen promptly in the case study organization. TOGAF [28] as an enterprise architecture framework, provides a widely accepted and increasingly applied techniques for enterprise architecture development. In TOGAF, the link between the strategy of the enterprise and the required business model is established and the high-level enterprise BPM process models are planned [28]. It is in the process of establishing a more robust EA practice.

Vara et al., [29] attempted to prevent common mistakes detected in practice such as the misconception in understanding the business process by system analysts, the lack of focus on system goals, and miscommunication between stakeholders and system analysts. Their proposed method is based on the goal analysis via BPMN and the MAP model. Map is a goal/strategy-driven approach to capture the purposes of a system and determine the strategies that can contribute to fulfilling these goals.

## V. Conclusion

In this paper, we had a survey on BORE method for requirements engineering, and the important lessons of BDD were introduced. Then we formed our combinational approach for requirements elicitation in order to develop business models. To do this, BORE method is used as the base of our work and it is enriched by the features of the BDD. It let us to model the more complex processes. One of the advantages of this method was exact requirements elicitation that adapted the customers' needs. Also, it prevented us from rework in the process modeling. The case study of the paper submission and publication system of journal corroborated the proficiency of this approach for modeling the complex systems with many sub-processes and complicated relationships. Also, it was a good experience of requirements engineering and modeling a real-world web-based application.

## References

[1] Nosrati, M., Karimi, R., Mohammadi, M., & Malekian, K. (2013). Internet Marketing or Modern Advertising! How? Why. International Journal of Economy, Management and Social Sciences, 2(3), 56-63.

[2] Khoshnampour, M., & Nosrati, M. (2011). An overview of E-commerce. World Applied Programming, 1(2), 94-99.

[3] B. Nuseibeh and S. Easterbrook, "Requirements engineering: a roadmap," in Proceedings of the conference on the future of Software engineering (ICSE'00), pp. 35-46, 2000.

[4] A. Przybyłek, "A Business-Oriented Approach to Requirements Elicitation," in Proceedings of the 9th International Conference on Evaluation of Novel Software Approaches to Software Engineering (ENASE'14), pp. 152-163, 2014.

[5] Faulk, S., 1997. Software Requirements: A Tutorial. In: Thayer, R., Dorfman, M. (Eds.): Software Requirements Engineering. IEEE Computer Society press.

[6] Maciaszek, L., 2005. Requirements analysis and system design. Addison-Wesley.

[7] Taylor-Cummings, A., 1998. Bridging the user-IS gap: a study of major information systems projects. Journal of Information Technology 13, pp. 29–54.

[8] Monsalve, C., April, A., Abran, A., 2010. Representing Unique Stakeholder Perspectives in BPM Notations. In: 8th ACIS International Conference on Software Engineering Research, Management and Applications, Montreal, Canada.

[9] Ravichandar, R., Arthur, J. D., Perez-Quinones, M., 2007. Pre-requirement specification traceability: Bridging the complexity gap through capabilities.

[10] Gotel, O., Finkelstein, A., 1994. An Analysis of the Requirements Traceability Problem. In: 1st International Conference on Requirements Engineering, Colorado Springs, CO.

[11] Nosrati, M., Hariri, M., & Shakarbeygi, A. (2013). Computers and internet: From a criminological view. International Journal of Economy, Management and Social Sciences, 2(4), 104-107.

[12] Nosrati, M., Karimi, R., Makekian, K., & Hariri, M. (2013). PayPal as the most loved payment system among merchants and buyers in online transactions. World Applied Programming, 3(9), 396-400.

[13] Nosrati, M., Nosrati, M., Karimi, R., & Karimi, R. (2016). Energy efficient and latency optimized media resource allocation. International Journal of Web Information Systems, 12(1), 2-17.

[14] Sadeghi, M., Mohammadi, M., Nosrati, M., & Malekian, K. (2013). The Role of Entrepreneurial Environments in University Students Entrepreneurial Intention. World Applied Programming, 3(8), 361-366.

[15] Nosrati, M., & Karimi, R. (2016). Investigating a benchmark cloud media resource allocation and optimization. World Applied Programming, 6(1), 5-9.

[16] Nosrati, M., Chalechale, A., & Karimi, R. (2015, June). Latency optimization for resource allocation in cloud computing system. In International Conference on Computational Science and Its Applications (pp. 355-366). Springer International Publishing.

[17] Nosrati, M., Karimi, R., & Hariri, M. (2013). General trends in multiplayer online games. World Applied Programming, 3(1), 1-4.

[18] Nosrati, M., Hanani, A., & Karimi, R. (2015). In proc. of Advanced Computing & Communication Technologies (ACCT), 2015 Fifth International Conference on. IEEE, 102-107.

[19] Nosrati, M., & Hariri, M. (2011). An Algorithm for Minimizing of Boolean Functions Based on Graph DS. World Applied Programming, 1(3), 209-214.

[20] Weske, M., Business Process Management - Concepts, Languages, Achitectures. 2007: Springer.

[21] T. Mitra, "Business-driven development," IBM developer works, http://www.ibm.com/developerworks/library/ws-bdd/index.html, 2005.

[22] H. Hiisilä, M. Kauppinen and S. Kujala, "Challenges of the customer organization's requirements engineering process in the outsourced environment - A case study," in Requirements Engineering: Foundation for Software Quality, pp. 214–229, 2015.

[23] I. Jacobson, G. Booch and J. Rumbaugh, The unified software development process, Addison-Wesley, 1999.

[24] Business Rules Group, "Defining Business Rules ~ What Are They Really?," 2000.

[25] M. Glinz, "On Non-Functional Requirements," in 15th IEEE International Requirements Engineering Conference, 21-26, 2007.

[26] L. Chung and J. C. Sampaio do Prado Leite, "On Non-Functional Requirements in Software Engineering," Chung, Lawrence, and Julio Cesar Sampaio do Prado Leite. "On non-functional requirements in Conceptual modeling: Foundations and applications, pp. 363-379, 2009.

[27] Štolfa, S., Vondrák, I., 2004. A Description of Business Process Modeling as a Tool for Definition of Requirements Specification. In: 12th International Conf. on Systems Integration, Prague, Czech Republic.

[28] The Open Group, "The Open Group Architecture Framework (TOGAF) Version 9.1," The Open Group, 2011.

[29] Vara, J.L., Sánchez, J., Pastor, Ó., 2008. Business Process Modelling and Purpose Analysis for Requirements Analysis of Information Systems. In: 20th international conference on Advanced Information Systems Engineering, Montpellier, France.